\documentclass[12pt]{article}
\usepackage{latexsym}
\usepackage{amsmath}
\usepackage{amssymb}
\usepackage{epsfig}
\usepackage{epic}

\numberwithin{equation}{section}
\numberwithin{figure}{subsection}

\begin{document}

\def\k{\kappa}
\def\half{\fraction{1}{2}}
\def\fraction#1#2{ { \scriptstyle \frac{#1}{#2} }}
\def\der#1{\frac{\partial}{\partial #1}}
\def\d{\partial}
\def\p2{\fraction{\pi}{2}}
\def\s{\sigma}
\def\Id{|I\rangle}
\def\Tr{\mathrm{Tr}}
\def\z{\xi}
\def\eq#1{eq.(\ref{eq:#1})}
\def\Eq#1{Eq.(\ref{eq:#1})}
\def\L{\mathcal{L}}
\def\B{\mathcal{B}}

\begin{titlepage}
\rightline{\today}

\begin{center}
\vskip 1.0cm
{\Large \bf{Marginal Solutions for the Superstring} }
\\
\vskip 1.0cm

{\large Theodore Erler}
\vskip 1.0cm
{\it {Harish-Chandra Research Institute} \\ 
{Chhatnag Road, Jhunsi, Allahabad 211019, India}}
\\ E-mail:terler@mri.ernet.in \\

\vskip 1.0cm
{\bf Abstract}
\end{center}

We construct a class of analytic solutions of WZW-type open superstring
field theory describing marginal deformations of a reference D-brane 
background. The deformations we consider are generated by on-shell 
vertex operators with vanishing operator products. The superstring solution
exhibits an intriguing duality with the corresponding marginal solution of 
the {\it bosonic} string. In particular, the superstring problem is ``dual''
to the problem of re-expressing the bosonic marginal solution in pure gauge
form. This represents the first nonsingular analytic solution of open 
superstring field theory.

\noindent

\medskip

\end{titlepage}

\newpage

\baselineskip=18pt

\tableofcontents

\section{Introduction}

Following the breakthrough analytic solution of Schnabl\cite{Schnabl}, our 
analytic understanding of open string field theory (OSFT) has seen remarkable 
progress\cite{Okawa,Fuchs,RZ,cohomology,amplitudes,RZO,Erler}. 
So far most work has focused on the open bosonic string, but clearly it is 
also important to consider the superstring. This is not just because 
superstrings are ultimately the theory of interest, but because there are 
important physical questions, especially the holographic encryption of closed 
string physics in OSFT, which may be difficult to decipher in the bosonic 
case\cite{Taylor}.

Ideally, the first goal should be to find an analytic solution of
superstring field theory\footnote{In this paper we will work with the Berkovits
WZW-type superstring field theory\cite{Berkovits}.} on a non-BPS brane 
describing the endpoint of tachyon condensation, i.e. the closed string 
vacuum. However, the construction of this solution is will likely be 
subtle---indeed, Schnabl's solution for the bosonic vacuum is very close to 
being pure gauge\cite{Schnabl,Okawa}. Thus, it may be useful to consider a 
simpler problem first: constructing solutions describing {\it marginal 
deformations} of a (non)BPS D-brane. Marginal deformations correspond to a 
one-parameter family of open string backgrounds obtained by adding a conformal
boundary interaction to the worldsheet action---for example, turning on a 
Wilson line on a brane by adding the boundary term 
$A_\mu\int_{\d\Sigma}dt \d X^\mu(t)$ to the worldsheet action. Such backgrounds
were studied numerically for the bosonic string in ref.\cite{marginal} and
for the superstring in ref.\cite{super_marginal}. Recently, 
Schnabl\cite{Schnabl2} and Kiermaier {\it et al}\cite{RZOK} found 
{\it analytic} solutions for marginal deformations in bosonic 
OSFT\footnote{For previous efforts to construct such solutions analytically in 
bosonic and super OSFT, see refs.\cite{Id,super_id}.}. The 
solutions bear striking resemblance to Schnabl's vacuum solution, but are 
simpler in the sense that they are manifestly nontrivial and can be 
constructed systematically with a judicious choice of gauge. 

In this note, we construct solutions of super OSFT describing marginal 
deformations generated by on-shell vertex operators with {\it vanishing} 
operator products (in either the $0$ or $-1$ picture). As was found in
ref.\cite{Schnabl2,RZOK} such deformations are technically simpler since
they allow for solutions in Schnabl's gauge, $\mathcal{B}_0\Phi=0$---though
probably more general marginal solutions can be obtained once the 
analogous problem is understood for the bosonic string, either
by adding counterterms as described in ref.\cite{RZOK} or by employing a 
``pseudo-Schnabl gauge'' as suggested in ref.\cite{Schnabl2}. The 
superstring solution exhibits a remarkable duality with its bosonic 
counterpart: it formally represents a re-expression of the bosonic solution in 
pure gauge form. It would be very interesting if this duality generalized to 
other solutions. 

This paper is organized as follows. In section \ref{sec:Review} we briefly 
review the bosonic marginal solution in the split string 
formalism\cite{Okawa,Erler,SSF}, which we will prove convenient for many 
computations. In section \ref{sec:Solution} we consider the superstring, 
motivating the solution as analogous to constructing an 
explicit pure gauge form for the bosonic marginal solution. This
strategy quickly gives a very simple expression for the complete analytic 
solution of super OSFT. In section \ref{sec:pure_gauge} we consider the 
dual problem: finding a pure gauge expression for the bosonic marginal 
deformation describing a constant, light-like gauge field on a non-compact 
brane. Though quite analogous to the superstring, this problem is slightly 
more complex. Nevertheless we are able to find an analytic solution. We end 
with some conclusions.

While this note was in preparation, we learned of the independent solution
by Yuji Okawa\cite{Yuji}. His paper should appear concurrently.

\section{Bosonic Solution}
\label{sec:Review}

Let us begin by reviewing the bosonic marginal solution\cite{Schnabl2,RZOK} 
in the language of the split string formalism\cite{Okawa,Erler,SSF}, 
which is a useful shorthand for many calculations. The first 
step in this approach is to find a subalgebra of the open string star 
algebra, closed under the action of the BRST operator, in which we hope to 
find an analytic solution. For the bosonic marginal solution the
subalgebra is generated by three string fields $K,B$ and $J$:
\begin{eqnarray}K &=& \mathrm{Grassmann\ even,\ gh}\#= 0\nonumber\\ 
B &=& \mathrm{Grassmann\ odd,\ gh}\#= -1\nonumber\\
J &=& \mathrm{Grassmann\ odd,\ gh}\#= 1\end{eqnarray}
satisfying the identities,
\begin{equation}[K,B]= 0 \ \ \ \ B^2 = J^2 = 0\label{eq:bosonic_id1}
\end{equation}
and
\begin{equation}dK = 0\ \ \ dJ=0\ \ \ dB=K\label{eq:bosonic_id2}\end{equation}
where $d=Q_B$ is the BRST operator and the products above are open string
star products (we will mostly omit the $*$ in this paper). The relevant 
explicit definitions of $K,B,J$ are\footnote{We may generalize the construction
by considering other projector frames\cite{RZ,RZO,Erler} or by allowing
the field $F$ in \eq{bosonicSOL} to be an arbitrary function of 
$K$\cite{Okawa,Erler}. Such generalizations do not add much to the current 
discussion so we will stick with the definitions presented here.},
\begin{eqnarray}
K &=& -\frac{\pi}{2}(K_1)_L\Id\ \ \ \ \ K_1 = L_1+L_{-1}\nonumber\\ 
B &=& -\frac{\pi}{2}(B_1)_L\Id\ \ \ \ \ B_1 = b_1+b_{-1}\nonumber\\ 
J &=& J(1)\Id
\label{eq:KBJ}\end{eqnarray}
where $\Id$ is the identity string field and the subscript $L$ denotes taking
the left half of the corresponding charge\footnote{``Left'' means integrating 
the current counter-clockwise on the positive half of the unit circle. This 
convention differs by a sign from ref.\cite{Erler} but agrees with 
ref.\cite{RZ}.}. The operator $J(z)$
is a dimension zero primary generating the marginal trajectory. It takes
the form,
\begin{equation}J(z)=c\mathcal{O}(z)\end{equation}
where $\mathcal{O}$ is a dimension one matter primary with 
{\it nonsingular} OPE with itself. This is crucial for guaranteeing that 
the square of the field $J$ vanishes, as in \eq{bosonic_id1}. With these
preliminaries, the marginal solution for the bosonic string is:
\begin{equation}\Psi = 
\lambda FJ\frac{1}{1-\lambda B\frac{F^2-1}{K}J}F
\label{eq:bosonicSOL}\end{equation} where $\lambda$ parameterizes the marginal
trajectory and $F = e^{K/2} = \Omega^{1/2}$ is the square root of the 
$SL(2,\mathbb{R})$ vacuum (a wedge state). To linear order in $\lambda$ the
solution is,
\begin{equation}\Psi = \lambda FJF+...= \lambda J(0)|\Omega\rangle+...
\end{equation} 
which is the nontrivial element of the BRST cohomology generating the marginal
trajectory. 

Let us prove that eq.\ref{eq:bosonicSOL} satisfies the equations of motion.
Using the identities Eqs.(\ref{eq:bosonic_id1},\ref{eq:bosonic_id2}),
\begin{eqnarray}d\Psi &=&  
-\lambda FJd\left(\frac{1}{1-\lambda B\frac{F^2-1}{K}J}\right)F\nonumber\\
&=& -\lambda FJ\frac{1}{1-\lambda B\frac{F^2-1}{K}J}
d\left(\lambda B\frac{F^2-1}{K} J\right)\frac{1}{1-\lambda B\frac{F^2-1}{K}J}F
\nonumber\\
&=& -\lambda^2 FJ\frac{1}{1-\lambda B\frac{F^2-1}{K}J}
(F^2-1) J \frac{1}{1-\lambda B\frac{F^2-1}{K}J}F
\end{eqnarray}
Notice the $(F^2-1)J$ factor in the middle. Since $J^2=0$, the $-1)J$ term 
vanishes when multiplied with the $J$s to the left---thus the necessity of 
marginal operators with nonsingular OPE. This leaves,
\begin{equation}d\Psi = -\lambda^2 FJ\frac{1}{1-\lambda B\frac{F^2-1}{K}J}
F^2 J \frac{1}{1-\lambda B\frac{F^2-1}{K}J}F = -\Psi^2\end{equation}
i.e. the bosonic equations of motion are satisfied.

The solution has a power series expansion in $\lambda$:
\begin{equation}\Psi = \sum_{n=1}^\infty\lambda^n\Psi_n\end{equation}
where,
\begin{equation}\Psi_n = FJ\left(B\frac{F^2-1}{K}J\right)^{n-1}F\end{equation}
To make contact with the expressions of refs.\cite{Schnabl2,RZOK}, note the
relation, 
\begin{equation}\frac{F^2-1}{K}=\int_0^1 dt\Omega^t\label{eq:int_rep}
\end{equation}
To prove this, recall $\Omega^t = e^{tK}$ and calculate\footnote{Note that,
in general, the inverse of $K$ is not well defined. However, when operating on
$F^2-1$ it is. This is why we cannot simply use $F^2/K$ in the solution 
in place of $\frac{F^2-1}{K}$, which would naively give a solution even for
marginal operators with singular OPEs.},
\begin{equation}K\int_0^1dt \Omega^t = \int_0^1 dt \frac{d}{dt}e^{tK}
= e^K-1 = F^2-1
\end{equation}
Using this and the mapping between the split string notation and conformal 
field theory described in ref.\cite{Erler}, the $\Psi_n$s can be written as
CFT correlators on the cylinder:
\begin{equation}\langle \Psi_n,\chi\rangle=
\int_0^1 dt_1...\int_0^1 dt_{n-1}\left\langle 
J(t_{n-1} + ...+ t_1 + 1)B... J(t_1+1)BJ(1)\,\,
f_\mathcal{S}\circ\chi(0)\right\rangle_{C_{t_{n-1}+...+t_1+2}}
\end{equation}
where $f_\mathcal{S}(z) = \frac{2}{\pi}\tan^{-1}z$ is the sliver conformal
map, and in this context $B$ is the insertion 
$\int_{i\infty}^{-i\infty}\frac{dz}{2\pi i}b(z)$ to be integrated parallel to 
the axis of the cylinder in between the $J$ insertions on either side. This 
matches the expressions found in refs.\cite{Schnabl2,RZOK}. 

In passing, we mention that this solution was originally constructed 
systematically by using the equations of motion to recursively determine the 
$\Psi_n$s in Schnabl gauge. If desired, it is also possible to perform such 
calculations in split string language; we offer some sample calculations in 
appendix \ref{ap:BL}.

\section{Superstring Solution}
\label{sec:Solution}

Let us now consider the superstring. The marginal deformation is generated 
by a $-1$ picture vertex operator,
\begin{equation}e^{-\phi}c\mathcal{O}(z)\end{equation}
where $\mathcal{O}(z)$ is a dimension $\frac{1}{2}$ superconformal matter 
primary. We will use Berkovits's WZW-type superstring field 
theory\cite{Berkovits}\footnote{See refs.\cite{Berkovits3,Ohmori,deSmet} for 
nice reviews.}, in which case the string field is given by multiplying
the $-1$ picture vertex operator by the $\xi$ ghost:
\begin{equation}X(z) = \xi e^{-\phi}c\mathcal{O}(z)\label{eq:X}
\end{equation} This corresponds to a solution of the linearized Berkovits 
equations of motion,
\begin{equation}\eta_0 Q_B\left(\lambda X(0)|\Omega\rangle\right)=0
\end{equation}
since $\eta_0$ eats the $\xi$ and the $-1$ picture vertex operator is in 
the BRST cohomology. We will also find it useful to consider the $0$ picture 
vertex operator,
\begin{equation}J(z) = Q_B\cdot X(z)=c G_{-1/2}\cdot\mathcal{O}(z) 
- e^\phi\eta\mathcal{O}(z)
\label{eq:J}\end{equation} A complimentary way of seeing the linearized 
equations of motion are satisfied is to note that $J(z)$ is in the small 
Hilbert space. As with the bosonic string, it is very helpful to assume that 
$X(z)$ and $J(z)$ have vanishing OPEs:
\begin{equation}\lim_{z\to w}J(z)X(w) = \lim_{z\to w}J(z)J(w) = \lim_{z\to w}
X(z)X(w) = 0\end{equation}
We mention two examples of such deformations. The simplest is the light-like 
Wilson line $\mathcal{O}(z)=\psi^+(z)$ ($\alpha'=1$), where
\begin{eqnarray}
X(z) &=& \xi e^{-\phi}c\psi^+(z)\nonumber\\
J(z) &=& i\sqrt{2}c\d X^+(z)-e^\phi\eta\psi^+(z)\end{eqnarray}
There is also a ``rolling tachyon'' marginal deformation\cite{Sen} 
$\mathcal{O}(z) = \sigma_1e^{X^0/\sqrt{2}}(z)$ on a non-BPS brane. The 
corresponding vertex operators are,
\begin{eqnarray}
X(z) &=& \sigma_1 \xi e^{-\phi}c e^{X^0/\sqrt{2}}(z)\nonumber\\
J(z) &=& \sigma_2(c\psi^0-ie^\phi\eta) e^{X^0/\sqrt{2}}(z)
\end{eqnarray}
The Pauli matrices $\sigma_1,\sigma_2,\sigma_3$ are ``internal'' Chan-Paton 
factors\cite{Berkovits2,BSZ}, necessary to accommodate non-BPS GSO$(-)$ states 
into the Berkovits framework. Though we will not write it explicitly, in this 
context it is important to remember that the BRST operator and the eta zero 
mode are carrying a factor of $\sigma_3$ (thus the presence 
$i\sigma_2 = \sigma_3\sigma_1$ in the above expression for $J$). We mention 
that both $X(0)|\Omega\rangle$ and $J(0)|\Omega\rangle$ are in Schnabl gauge
and annihilated by $\L_0$.

Let us describe the subalgebra relevant for finding the marginal solution. 
It consists of the products of four string fields, $K,B,X,J$\footnote{Note that
for for a GSO$(-)$ deformation the Grassmann assignments of $X,J$ are opposite.
Still, as far as the solution is concerned $X$ is even and $J$ is odd 
because $Q_B,\eta_0$ carry a $\sigma_3$ which anticommutes with the internal
Chan-Paton matrices of the vertex operators.}:
\begin{eqnarray}K &=& \mathrm{Grassmann\ even,\ gh}\#= 0
\nonumber\\ 
B &=& \mathrm{Grassmann\ odd,\ gh}\#= -1\nonumber\\
X &=& \mathrm{Grassmann\ even,\ gh}\#= 0\nonumber\\
J &=& \mathrm{Grassmann\ odd,\ gh}\#= 1\label{eq:super_fields}
\end{eqnarray}
All four of these have vanishing picture number. $K$ and $B$ are the same
fields encountered earlier in \eq{KBJ}; $X$ and $J$ are defined,
\begin{equation}X = X(1)|I\rangle\ \ \ \ \ J=J(1)|I\rangle\end{equation}
with $X(z),J(z)$ as in Eqs.(\ref{eq:X},\ref{eq:J}). We have the identities,
\begin{equation}[K,B] = 0\ \ \ B^2=0\ \ \ X^2=J^2=XJ=JX=0 \label{eq:susy_id1}
\end{equation}
where the third set follows because the corresponding vertex operators have 
vanishing OPEs. The algebra is closed under the action of the BRST operator:
\begin{eqnarray}dB&=&K\ \ \ \ \ dK=0\nonumber\\
dX &=& J\ \ \ \ \ dJ=0\label{eq:susy_id2}\end{eqnarray}
Note that the eta zero mode $\bar{d}\equiv\eta_0$ annihilates $K,B$ and $J$, 
\begin{equation}\bar{d}K = \bar{d}B = \bar{d}J =0\end{equation}
since they live in the small Hilbert space. However, it does {\it not}
annihilate $X$, and the algebra is not closed under $\bar{d}$. Though
it is not {\it a priori} obvious that the $K,B,X,J$ algebra is rich enough 
to encapsulate the marginal solution, we will quickly see that it is.

We seek a one parameter family of solutions of the super OSFT equations of 
motion,
\begin{equation}\bar{d}\left(e^{-\Phi}de^\Phi\right)=0\label{eq:EOM}
\end{equation}
where $\Phi$ is a Grassmann even, ghost and picture number zero string
field which to linear order in the marginal parameter takes the form,
\begin{equation}\Phi = \lambda FXF+... \end{equation}
There are many strategies one could take to solve this equation, but
before describing our particular approach it is worth mentioning the 
``obvious'' method: fixing $\Phi$ in Schnabl gauge and attempting a 
perturbative solution, as in refs.\cite{Schnabl2,RZOK}:
\begin{equation}\Phi = \sum_{n=1}^\infty \lambda^n\Phi_n
\ \ \ \ \ \Phi_1 = FXF\label{eq:pertPHI}
\end{equation}
At second order\footnote{Explicitly, if we plug \eq{pertPHI} into the equations
of motion, we find a recursive set of equations of the form $\bar{d}d\Phi_n = 
\bar{d}\mathcal{F}_{n-1}[\Phi]$, where $\mathcal{F}_{n-1}[\Phi]$ depends 
on $\Phi_1,...,\Phi_{n-1}$. The Schnabl gauge solution is obtained by writing
$\Phi_n = \frac{\mathcal{B}_0}{\mathcal{L}_0}\mathcal{F}_{n-1}[\Phi]$.}, 
the Schnabl gauge solution is actually fairly simple:
\begin{equation}\Phi_2 = \frac{1}{2!}\left[FXB\frac{F^2-1}{K}JF
+FJB\frac{F^2-1}{K}XF\right]\label{eq:Phi_2}\end{equation} and seems quite
similar to the bosonic solution. At third order, however, we found an extremely
complicated expression (though still within the $K,B,X,J$ subalgebra). It seems
doubtful that a closed form solution for $\Phi$ in Schnabl gauge can
be obtained.

Since the Schnabl gauge construction appears complicated, we are lead to 
consider another approach. To motivate our particular strategy, we make two
observations: First, the combination $e^{-\Phi}de^\Phi$ which enters the 
superstring equations of motion {\it also} happens to be a pure gauge 
configuration from the perspective of bosonic OSFT. Second, there is a basic 
similarity between the $K,B,J$ algebra for the bosonic marginal solution and 
the $K,B,J,X$ algebra for the superstring. The main difference of course is 
the presence of $X$ for the superstring, whose BRST variation gives $J$. If 
such a field were present for the bosonic string, the bosonic marginal solution
would be pure gauge because $J$ would be trivial in the BRST cohomology. 
With this motivation, we are lead to consider the equation
\begin{equation}e^{-\Phi}de^\Phi = \lambda 
FJ\frac{1}{1-\lambda B\frac{F^2-1}{K}J}F \label{eq:EOM2}\end{equation}
From the bosonic string perspective, this equation represents an expression of
the bosonic marginal solution in a form which is pure gauge. From the 
superstring perspective, this is a partially gauge fixed form of the equations
of motion, since the expression on the right hand side is in the small 
Hilbert space.

Let us now solve this equation. It will turn out to be simpler to solve for 
the group element $g=e^\Phi$; we make a perturbative ansatz,
\begin{equation}g=e^\Phi = 1+\sum_{n=1}^\infty \lambda^n g_n\ \ \ \ \ 
\ \ \ \ \ g_1=\Phi_1=FXF\end{equation}
Expanding out \eq{EOM2} to second order gives,
\begin{eqnarray}dg_2 &=& FJB\frac{F^2-1}{K}JF +g_1dg_1\nonumber\\
&=&  FJB\frac{F^2-1}{K}JF + FX F^2 JF \end{eqnarray}
As it turns out, this equation is solved by the second order Schnabl gauge 
solution \eq{Phi_2}:
\begin{equation}g_2=\Phi_2+\frac{1}{2}\Phi_1^2 = 
\frac{1}{2!}\left[FXB\frac{F^2-1}{K}JF
+FJB\frac{F^2-1}{K}XF+FXF^2XF\right]\end{equation}
but there is a simpler solution:
\begin{equation}g_2 = FXB\frac{F^2-1}{K}JF\end{equation}
Using this form of $g_2$ we can proceed to third order---remarkably, the 
solution is practically just as simple:
\begin{equation}g_3 = FX\left(B\frac{F^2-1}{K}J\right)^2 F\end{equation}
This leads to an ansatz for the full solution:
\begin{equation}e^\Phi = 1+\lambda FX\frac{1}{1-\lambda B\frac{F^2-1}{K}J}F
\label{eq:solution}
\end{equation}
To check this, calculate:
\begin{eqnarray}de^\Phi &=& \lambda FJ\frac{1}{1-\lambda B\frac{F^2-1}{K}J}F
+\lambda FXd\left(\frac{1}{1-\lambda B\frac{F^2-1}{K}J}\right)F\nonumber\\
&=& \lambda FJ\frac{1}{1-\lambda B\frac{F^2-1}{K}J}F 
+\lambda FX\frac{1}{1-\lambda B\frac{F^2-1}{K}J}d\left(\lambda 
B\frac{F^2-1}{K}J\right)
\frac{1}{1-\lambda B\frac{F^2-1}{K}J}F\nonumber\\
&=& \lambda FJ\frac{1}{1-\lambda B\frac{F^2-1}{K}J}F 
+\lambda^2 FX\frac{1}{1-\lambda B\frac{F^2-1}{K}J}F^2 J
\frac{1}{1-\lambda B\frac{F^2-1}{K}J}F\nonumber\\
&=&\left(1+\lambda FX\frac{1}{1-\lambda B\frac{F^2-1}{K}J}F \right)
\lambda FJ\frac{1}{1-\lambda B\frac{F^2-1}{K}J}F \nonumber\\
&=& e^{\Phi}\lambda FJ\frac{1}{1-\lambda B\frac{F^2-1}{K}J}F 
\end{eqnarray}
Therefore, \eq{solution} is indeed a complete solution to the super OSFT 
equations of motion! Note, however, that it is not quite a solution to the 
pure gauge problem of the bosonic string. In particular, in step three we 
needed to assume $XJ=0$---something we would not expect to hold in the
bosonic context. We will give the solution to the bosonic problem in the next
section. 

Let us make a few comments about this solution. First, though the string
field $\Phi$ itself is not in Schnabl gauge, the nontrivial part of the
group element $e^\Phi$ {\it is}---this is not difficult to see, but we offer
one explanation in appendix \ref{ap:BL}. The second comment is related to the
string field reality condition. In super OSFT, the natural reality 
condition is that $\Phi$ should be ``imaginary'' in the following sense:
\begin{equation}\langle \Phi,\chi\rangle = -\langle \Phi|\chi\rangle
\end{equation} where $\langle \Phi|$ is the Hermitian dual of $|\Phi\rangle$ 
and $\chi$ is any test state. In split string notation we can write
this,
\begin{equation}\Phi^\dag = -\Phi\end{equation} 
where $\dag$ is an anti-involution on the star algebra, formally completely
analogous to Hermitian conjugation of operators. With this reality condition,
the group element should be unitary:
$$g^{\dag}=g^{-1}$$
Using,
\begin{equation}K^\dag=K\ \ \ B^\dag=B\ \ \ J^\dag=J\ \ \ X^\dag = -X
\end{equation}
it is not difficult to see that the analytic solution $e^\Phi$ 
is {\it not} unitary\footnote{By contrast, the Schnabl gauge construction 
automatically gives an imaginary $\Phi$ and unitary $e^\Phi$.}. However, 
it is possible to obtain a unitary solution by a simple gauge transformation 
of \eq{solution}; we explain details in appendix \ref{ap:real_sol}.

Let us take the opportunity to express the solution in a few other forms which
may be more convenient for explicit computations. Following the usual 
prescription we may express the $g_n$s as correlation functions 
on the cylinder:
\begin{eqnarray}\langle g_n,\chi\rangle &=&
\int_0^1 dt_1...\int_0^1 dt_{n-1}\left\langle 
X(t_{n-1} + ...+ t_1 + 1)BJ(t_{n-2}+..+t_1+1)...BJ(1)\,\,
f_\mathcal{S}\circ\chi(0)\right\rangle_{C_{t_{n-1}+...+t_1+2}}\nonumber\\
&=&(-1)^n\int_0^1 dt_1...\int_0^1 dt_{n-1}\left\langle 
X(L+1)[\mathcal{O}'(\ell_{n-2}+1)...\mathcal{O}'(\ell_1+1)]BJ(1)\,\,
f_\mathcal{S}\circ\chi(0)\right\rangle_{C_{L+2}}
\end{eqnarray}
In the second line we manipulated the multiple $B$ insertions, simplifying the 
vertex operators and obtaining a single $B$ insertion to the right; we
introduced the length parameters\cite{RZOK}:
\begin{equation}\ell_i = \sum_{k=1}^i t_k\ \ \ \ \ \ 
L=\ell_{n-1}\end{equation} and defined 
$\mathcal{O}'(z) = G_{-\half}\cdot\mathcal{O}(z)$ 
(times a $\sigma_3$ for GSO$(-)$ deformations). We may also express the 
solution in the operator formalism of Schnabl\cite{Schnabl}:
\begin{eqnarray}|g_n\rangle &=& \frac{(-1)^{n\mathcal{O}+1}}{2}
\int_0^1 dt_1...\int_0^1dt_{n-1}\hat{U}_{L+2}\,f_\mathcal{S}^{-1}\circ
(\xi e^{-\phi}\mathcal{O}(L/2))\tilde{\mathcal{O}}'(y_{n-2})...
\tilde{\mathcal{O}}'(y_1)\nonumber\\
&\ &\ \ \ \times\left(\tilde{\mathcal{O}}'(-\fraction{L}{2})
[\mathcal{B}^+\tilde{c}(\fraction{L}{2})\tilde{c}(-\fraction{L}{2})
-\tilde{c}(\fraction{L}{2})-\tilde{c}(-\fraction{L}{2})]
+f_\mathcal{S}^{-1}\circ(\eta e^\phi\mathcal{O}(-\fraction{L}{2}))
[\mathcal{B}^+\tilde{c}(\fraction{L}{2})+1]\right)|\Omega\rangle
\nonumber\\ \end{eqnarray} where $y_i=\ell_i-L/2$ and\cite{amplitudes} 
$\hat{U}_r=\left(\frac{2}{r}\right)^{\L_0^*}\left(\frac{2}{r}\right)^{\L_0}$.
Also we have used $f_\mathcal{S}^{-1}$ to define the tilde to hide some
factors of $\frac{\pi}{2}$. The expression is somewhat more complicated than
the bosonic solution since the vertex operator $J(z)$ has a piece 
without a $c$ ghost, so in the $bc$ CFT the solution has a component not 
proportional to Schnabl's $\psi_n$\cite{Schnabl}.

\section{Pure Gauge for Bosonic Solution}
\label{sec:pure_gauge}

In the last section, we found a solution for the superstring by analogy with
the pure gauge problem of the bosonic string; but we did not solve the latter.
The scenario we have in mind is a constant, lightlike gauge field on a 
non-compact D-brane. Since there is no flux and no way to wind a Wilson loop, 
such a field configuration should be pure gauge. From the string field
theory viewpoint, this is reflected by the fact that the marginal vertex 
operator becomes BRST trivial in the noncompact limit,
\begin{equation}ic\d X^+(z) = 
Q_B\cdot 2i X^+(z)\end{equation}
Of course, on a compact manifold the operator $X^+(z)$ is not globally 
defined so the marginal deformation is nontrivial.

Translating to split string language, we consider an algebra generated by 
four fields $K,B,X,J$, where $K,B$ are defined as before and,
\begin{equation}X = 2iX^+(1)\Id\ \ \ \ 
J=i c\d X^+(1)\Id\end{equation}
These have the same Grassmann and ghost number assignments as 
\eq{super_fields}. We have the algebraic relations,
\begin{equation}[K,B] = 0\ \ \ \ \ \ B^2=0\ \ \ \ \ J^2=0
\ \ \ \ \ [X,J]=0\end{equation}
Note the difference from the superstring case: the products of $X$ with 
itself and with $J$, though well defined (the OPEs are nonsingular), are 
nonvanishing. However, we still have
\begin{eqnarray}dB&=&K\ \ \ \ \ dK=0\nonumber\\
dX &=& J\ \ \ \ \ dJ=0\end{eqnarray}
with the second set implying that $J$ is trivial in the BRST cohomology. 

We now want to solve \eq{EOM2} assuming this slightly more general set of 
algebraic relations. Playing around a little bit, the solution we found
is,
\begin{equation}e^\Lambda = 1+\lambda F u_\lambda(X)
\frac{1}{1-\lambda B\frac{F^2-1}{K}J}F\end{equation}
where,
\begin{equation}u_\lambda(X) = \frac{e^{\lambda X}-1}{\lambda}\end{equation}
The relevant identity satisfied by this particular combination is,
\begin{equation}du_\lambda = J(\lambda u_\lambda +1)\end{equation}
Let us prove that this gives a pure gauge expression for the bosonic
marginal solution:
\begin{eqnarray}de^\Lambda &=& \lambda Fdu_\lambda
\frac{1}{1-\lambda B\frac{F^2-1}{K}J}F
+\lambda Fu_\lambda\frac{1}{1-\lambda B\frac{F^2-1}{K}J}d\left(\lambda 
B\frac{F^2-1}{K}J\right)
\frac{1}{1-\lambda B\frac{F^2-1}{K}J}F\nonumber\\\nonumber\\
&=& \lambda FJ(\lambda u_\lambda +1)\frac{1}{1-\lambda B\frac{F^2-1}{K}J}F 
+\lambda^2 Fu_\lambda\frac{1}{1-\lambda B\frac{F^2-1}{K}J}(F^2-1)J
\frac{1}{1-\lambda B\frac{F^2-1}{K}J}F\nonumber
\end{eqnarray}
Now we come to the critical difference from the superstring. Note the 
$-1)J$ piece in the middle of the second term. Before it vanished when
multiplied by $X,J$ to the left. This time it contributes because $XJ\neq 0$;
still, the $J$s in the denominator of the factor to the left get killed because
$J^2=0$. Thus we have,
\begin{eqnarray}de^\Lambda &=& 
\lambda FJ(\lambda u_\lambda +1)\frac{1}{1-\lambda B\frac{F^2-1}{K}J}F 
+\lambda^2 Fu_\lambda\frac{1}{1-\lambda B\frac{F^2-1}{K}J}F^2J
\frac{1}{1-\lambda B\frac{F^2-1}{K}J}F\nonumber\\ &\ &\ \ \ \ \ \ \ \ \ \ \ \ 
\ \ \ \ \ \ \ \ \ \ \ \ \ \ \ \ \ \ \ \ \ \ \ \ \ \ \ \ \ \ \ \ \ \ \ \ \ \ \ 
-\lambda^2 Fu_\lambda J
\frac{1}{1-\lambda B\frac{F^2-1}{K}J}F\end{eqnarray}
where the third term comes from the $-1)J$ piece. Note the cancellation. 
We get,
\begin{eqnarray}de^\Lambda &=& 
\lambda FJ\frac{1}{1-\lambda B\frac{F^2-1}{K}J}F 
+\lambda^2 Fu_\lambda\frac{1}{1-\lambda B\frac{F^2-1}{K}J}F^2J
\frac{1}{1-\lambda B\frac{F^2-1}{K}J}F\nonumber\\
&=& \left(1+\lambda Fu_\lambda\frac{1}{1-\lambda B\frac{F^2-1}{K}J}F\right)
\lambda FJ\frac{1}{1-\lambda B\frac{F^2-1}{K}J}F\nonumber\\
&=& e^\Lambda \lambda FJ\frac{1}{1-\lambda B\frac{F^2-1}{K}J}F\end{eqnarray}
thus we have a pure gauge expression for the marginal solution.

To further emphasize the duality with the superstring, note that for the 
pure gauge problem the role of the eta zero mode is played by the lightcone
derivative:
\begin{equation}\bar{d}\sim\frac{d}{dx^+}\end{equation}
In particular we have solved the equation,
\begin{equation}\frac{d}{dx^+}\left(e^{-\Lambda} de^\Lambda\right) = 0
\end{equation}
Though there are many pure gauge trajectories generated by $FXF$, only a 
trajectory which in addition satisfies this equation will be a well-defined,
nontrivial solution once spacetime is compactified.

\section{Conclusion}
\label{sec:Conclusion}

In this note, we have constructed analytic solutions of open superstring field
theory describing marginal deformations generated by vertex operators with 
vanishing operator products. We have not attempted to perform any detailed 
calculations with these solutions, though such calculations are certainly 
possible. The really important questions about marginal solutions---such
as mapping out the relation between CFT and OSFT marginal parameters, 
obtaining analytic solutions for vertex operators with singular OPEs, or 
proving Sen's rolling tachyon conjectures\cite{Sen}---require more work
even for the bosonic string. Hopefully progress will translate directly to the
superstring.

For us, the main motivation was the hope that marginal solutions could give
us a hint about how to construct the vacuum for the open superstring.
Indeed, for the bosonic string the marginal and vacuum solutions
are closely related: To get the vacuum solution (up to the $\psi_N$ piece), 
one simply replaces $J$ with $d(Bc)=cKBc$ and takes the limit 
$\lambda\to\infty$\footnote{The $\lambda$ used here and the $\lambda$ 
parameterizing the pure gauge solutions of Schnabl\cite{Schnabl} are 
related by $\lambda(Schnabl)=\frac{\lambda}{\lambda+1}$.}. Perhaps a similar
trick will work for the superstring. 

The author would like to thank A. Sen and D. Gross for conversations, and 
A. Bagchi for early collaboration. The author also thanks Y. Okawa for 
correspondence which motivated discovery of the unitary analytic solution 
presented in appendix \ref{ap:real_sol}. 
This work was supported in part by the 
National Science Foundation under Grant No.NSF PHY05-51164 and by the 
Department of Atomic Energy, Government of India.

\begin{appendix}
\section{$\B_0,\L_0$ with Split Strings}
\label{ap:BL}

In many analytic computations in OSFT it is useful to invoke the operators 
$\mathcal{B}_0,\mathcal{L}_0$ and their cousins\cite{Schnabl,RZ}. To avoid 
unnecessary transcriptions of notation, it is nice to accommodate these
types of operations in the split string formalism. 

We begin by defining the fields,
\begin{equation}\L = (\L_0)_L\Id\ \ \ \ 
\L^* = (\L_0^*)_L\Id\end{equation}
and their $b$-ghost counterparts $\B,\B^*$. We can split
the operators $\L_0,\L_0^*$ into left/right halves non-anomalously because 
the corresponding vector fields vanish at the midpoint\cite{RZ}. The fields
$\L,\L^*$ satisfy the familiar special projector algebra,
\begin{equation}[\L,\L^*]=\L+\L^*\label{eq:liealg}\end{equation}
Following ref.\cite{RZ} we may define even/odd combinations,
\begin{equation}\L^+ = \L+\L^* = -K\ \ \ \ \ \ \ \ \ \ 
\L^- = \L-\L^*\end{equation} where $K$ is the field introduced before. .
Note that we have, 
\begin{eqnarray}\L_0\cdot\Psi &=& \L\Psi+\Psi\L^*\nonumber\\
\B_0\cdot\Psi &=& \B\Psi+(-1)^\Psi\Psi\B^*\end{eqnarray}
We can use similar formulas to describe the many related operators introduced
in ref.\cite{RZ}

Let us now describe a few convenient facts. Let $J(z)$ be a vertex operator
for a state $J(0)|\Omega\rangle$ in Schnabl gauge, and let $J=J(1)\Id$ be its 
corresponding field. Then,
\begin{equation}[\B^-,J]=0\end{equation}
where $[,]$ is the graded commutator. A similar result $[\L^-,J]=0$ holds if
$J(0)|\Omega\rangle$ is killed by $\L_0$. We also have the useful formulas,
\begin{equation}\L F = \frac{1}{2} F\L^-\ \ \ \ \ \ \ \ 
F\L^* = -\frac{1}{2} \L^-F\ \ \ \ \ \ \ \ \ 
[\L^-,\Omega^\gamma] = 2\gamma K\Omega^\gamma\label{eq:sp_id}\end{equation}
The third equation is a special case of,
\begin{equation}[\L^-,G(K)]=2KG'(K)\label{eq:gen_id}\end{equation}
with similar formulas involving $\B,\B^*$. Of course, these equations are 
well-known consequences of the Lie algebra \eq{liealg}. 

As an application, let us prove the identity,
\begin{equation}\frac{\B_0}{\L_0}J_1(0)|\Omega\rangle*J_2(0)|\Omega\rangle = 
(-1)^{J_1}FJ_1B\frac{F^2-1}{K}J_2F\end{equation} 
where $J_1,J_2(0)|\Omega\rangle$ are
killed by $\B_0,\L_0$. This expression occurs when constructing the marginal
solution (bosonic or superstring) in Schnabl gauge. The direct approach is to
compute $\L_0^{-1}$ on the left hand side in split string notation; the 
resulting derivation is fairly reminiscent of ref.\cite{RZOK}. Instead, we will
multiply this equation by $\L_0$ and prove that both sides are equal. 
The left hand side gives,
\begin{eqnarray}\B_0\cdot FJ_1F^2J_2F &=& \B FJ_1F^2J_2F+(-1)^{J_1+J_2}
FJ_1F^2J_2F\B^*\nonumber\\
&=& \frac{1}{2}(-1)^{J_1}FJ_1[\B^-,F^2]J_2F\nonumber\\
&=& (-1)^{J_1}FJ_1 B F^2 J_2F
\end{eqnarray}
The right hand side gives,
\begin{eqnarray}\L_0\cdot FJ_1B\frac{F^2-1}{K}J_2F &=& 
\L FJ_1B\frac{F^2-1}{K}J_2F + FJ_1B\frac{F^2-1}{K}J_2F\L^*\nonumber\\
&=&\frac{1}{2}FJ_1\left[\L^-, B\frac{F^2-1}{K}\right]J_2F\nonumber\\
&=& FJ_1B\frac{F^2-1}{K}J_2F + \frac{1}{2}
FJ_1B\left[\L^-,\frac{F^2-1}{K}\right]J_2F
\end{eqnarray}
Focus on the commutator:
\begin{eqnarray}\left[\L^-,\frac{F^2-1}{K}\right] &=& [\L^-,F^2]\frac{1}{K} + 
(F^2-1)\left[\L^-,\frac{1}{K}\right]\nonumber\\
&=& 2 F^2 - 2\frac{F^2-1}{K}
\end{eqnarray}
where we used \eq{gen_id}. This computation is a somewhat formal because the 
inverse of $K$ is not generally well defined, but it can be checked using the
integral representation \eq{int_rep}. Plugging the commutator back in, 
the $\frac{F^2-1}{K}$ terms cancel and we are left with,
\begin{equation}\L_0\cdot FJ_1B\frac{F^2-1}{K}J_2F = FJ_1BF^2J_2F\end{equation}
which after multiplying by $(-1)^{J_1}$ establishes the result.

Before concluding, we mention that any state of the form,
\begin{equation}F J_1 BG_2(K) J_2\, ...\, BG_n(K)J_nF\end{equation}
with $[\B^-,J_i]=0$, is in Schnabl gauge. The proof follows at once upon 
noting,
\begin{equation}[\B^-,BG(K)] = -
2B^2G'(K) = 0\end{equation}
so the entire expression between the $F$s commutes with $\B^-$. This is 
one way of seeing that the nontrivial part of the group element $e^\Phi-1$
for the superstring solution is in Schnabl gauge.

\section{Unitary $e^\Phi$}
\label{ap:real_sol}

The analytic solution \eq{solution} is very simple, but it has the 
disadvantage of not satisfying the standard reality condition, i.e. $e^\Phi$
is not unitary and $\Phi$ is not imaginary. Presumably there is an infinite 
dimensional array of marginal solutions which do satisfy the reality condition,
and some may have analytic descriptions. In this appendix we give one 
construction which is particularly closely related to our solution 
\eq{solution}. For a very interesting and completely different solution, we
refer the reader to an upcoming paper by Okawa\cite{Yuji2}. 

Our strategy will be to find a finite gauge transformation of $g$ in 
\eq{solution} yielding a unitary solution. The transformation is, 
\begin{equation}U = Vg\label{eq:U}\end{equation}
where $V$ is some string field of the form,
\begin{equation}V = 1+ dv\label{eq:V_gauge}\end{equation}
with $v$ carrying ghost number $-1$. A little thought reveals a natural 
candidate for $V$:
\begin{equation}V = \frac{1}{\sqrt{gg^\dag}}\end{equation}
where $g^\dag$ is the conjugate of \eq{solution}:
\begin{equation}g^\dag = 1-\lambda F\frac{1}{1-J\lambda B\frac{F^2-1}{K}}XF
\end{equation}
and we use the Hermitian definition of the square root. Intuitively, this is
just taking the original solution and dividing by its ``norm.'' More 
explicitly, if we define,
\begin{eqnarray}gg^\dag &=& 1+T\nonumber\\
T &=& \lambda FX\frac{1}{1-\lambda B\frac{F^2-1}{K}J} F-
\lambda F\frac{1}{1-J\lambda B\frac{F^2-1}{K}}XF\nonumber\\ &\ &\ \ \ \ \ \ \ 
\ \ \ \ \ \ \ \ \ \ \ \ \ \ \ \ \ \ \ \ \ 
-\lambda^2 FX\frac{1}{1-\lambda B\frac{F^2-1}{K}J}
F^2\frac{1}{1-J\lambda B\frac{F^2-1}{K}}XF\end{eqnarray}
then the required gauge transformation is given by the formal sum,
\begin{equation}V =\frac{1}{\sqrt{gg^\dag}} = 
\sum_{n=0}^\infty \left(\begin{matrix}-1/2 \\ n\end{matrix}\right)T^n 
\end{equation}

This proposal must be subject to two consistency checks. First, of course, is
that the field $U$ is actually unitary. The proof is straightforward:
\begin{eqnarray}UU^\dag &=& \frac{1}{\sqrt{gg^\dag}}gg^\dag
\frac{1}{\sqrt{gg^\dag}} = gg^\dag\frac{1}{\sqrt{gg^\dag}}
\frac{1}{\sqrt{gg^\dag}} = 1\nonumber\\
U^\dag U &=& g^\dag\frac{1}{\sqrt{gg^\dag}}\frac{1}{\sqrt{gg^\dag}}g = 
g^\dag (g^\dag)^{-1}g^{-1}g = 1
\end{eqnarray}
The second check is that $V$ is a gauge transformation of the form
\eq{V_gauge}. This follows if the field $T$ is BRST exact, $T=du$, since then 
we can write (for example),
\begin{equation}V = 1+ d\left(\sum_{n=1}^\infty 
\left(\begin{matrix}-1/2 \\ n\end{matrix}\right)u T^{n-1}\right) \end{equation}
A little guesswork reveals the following BRST exact expression for $T$:
\begin{equation}T = d\left(\lambda^2 FX\frac{1}{1-\lambda B\frac{F^2-1}{K}J}
B\frac{F^2-1}{K}X F\right)\end{equation}
This establishes not only that $U$ is an analytic solution, but (perhaps more 
importantly) that the simpler expression $g$ is in the same gauge orbit with
a solution satisfying the physical reality condition. This leaves no question
as to the physical viability of our original analytic solution \eq{solution}.

As usual, the unitary solution $U$ can be defined explicitly in terms of 
cylinder correlators by expanding \eq{U} as a power series in $\lambda$.  
Unfortunately this is somewhat tedious because the implicit dependence on 
$\lambda$ in \eq{U} is complicated. As an expansion for the imaginary field
$\Phi$, the first two orders agree with the Schnabl gauge solution 
(as they must\footnote{The reality condition fixes the form of the second
order solution uniquely within the $K,B,J,X$ subalgebra.}), while at third
order we find:
\begin{eqnarray}\Phi_3 &=& \frac{1}{2}\left(FXB\frac{F^2-1}{K}JB
\frac{F^2-1}{K}JF+FJB\frac{F^2-1}{K}JB\frac{F^2-1}{K}XF\right)\nonumber\\
&\ &\ \ +\frac{1}{4}\left(FXF^2JB\frac{F^2-1}{K} + FJB\frac{F^2-1}{K}XF^2
\right)XF\nonumber\\ &\ &\ \ 
-\frac{1}{4}FX\left(B\frac{F^2-1}{K}JF^2XF+F^2XB\frac{F^2-1}{K}JF\right)
+\frac{1}{3}(FXF)^3
\end{eqnarray}
This expression is much simpler than the Schnabl gauge solution at third 
order, which involves intricate constrained and entangled integrals over 
moduli separating vertex operator insertions.

\end{appendix}

\end{document}